\documentclass[intlimits,twoside,a4paper]{article}

\usepackage[cp1251]{inputenc}

\usepackage[eqsecnum]{cmpj3}

\usepackage{bm}

\issue{2020}{23}{4}{43701}
\doinumber{10.5488/CMP.23.43701}
\title[Microscopic theory of high-temperature superconductivity...]%
{Microscopic theory of high-temperature superconductivity in\\
strongly correlated electronic systems %
\thanks{The paper is devoted to the 60-th
birthday of Professor A. Shvaika}}
\author[N.M. Plakida]{N.M. Plakida}
\address{
 Joint Institute for Nuclear Research, 141980 Dubna, Russia }
\date{Received April 11, 2020, in final form June 23, 2020}
\begin{document}

\maketitle

\begin{abstract}
A consistent microscopic theory of superconductivity for strongly correlated
electronic systems is presented. The Dyson equation for the normal and anomalous Green functions for the projected (Hubbard) electronic operators is derived.  To compare various mechanisms of pairing, the extended Hubbard model is considered  where the intersite Coulomb repulsion  and the electron-phonon interaction are taken into account. We obtain the $d$-wave pairing with high-$T_c$
induced by the strong kinematical interaction of electrons with spin
fluctuations, while the  Coulomb repulsion  and the electron-phonon interaction  are suppressed for the $d$-wave pairing. These results  support the spin-fluctuation mechanism of high-temperature superconductivity in cuprates previously proposed in phenomenological models.

\end{abstract}

\section{Introduction}
\label{sec:1}

Intensive experimental investigations  of high-temperature superconductivity (HTSC) in copper-oxides (cuprates) discovered  by Bednorz and M\"uller~\cite{Bednorz86} more than 30 years ago produced a detailed information concerning  unconventional physical properties of cuprates (see, e.g., reference~\cite{Plakida10}).  However, the corresponding theoretical studies
of  various microscopical models have not yet resulted in a commonly accepted theory of
HTSC. The main problem in the theoretical study is that the conventional Fermi-liquid approach fails  to  describe  the  electronic structure of the cuprates due to strong electron correlations~\cite{Fulde95}. They are Mott-Hubbard  (more accurately, charge-transfer) antiferromagnetic  (AF) insulators
where the conduction band splits into two Hubbard subbands: the filled singly occupied subband
and the empty  doubly occupied ones. They become poor conductors when doped by holes or electrons in the corresponding subbands. To describe such strongly correlated metal one has to use for the subbands the projected (Hubbard) electronic operators which are difficult to treat. Various theoretical methods were used to take into account the complicated non-Fermionic character of these operators, such as
numerical simulation for finite clusters, variational approach, diagram technique,
cluster approximations, slave boson (electron) representation, etc. (see references~\cite{Plakida10,Avella11}).

Recently, we have developed  a microscopic theory of spin excitations~\cite{Vladimirov09,Vladimirov11} and superconductivity~\cite{Plakida99,Plakida07, Plakida13,Plakida14,Plakida16} for cuprates. We employed  the equation of motion method for the thermodynamic Green functions \cite{Bogoliubov59,Zubarev60} (GFs) in terms of the Hubbard operators (HOs)   which enabled  us to take into account rigorously the non-Fermionic character of electronic operators. We emphasize that the commutation relations for HOs   result in the  specific kinematical interaction  which plays an important role in studying  strongly correlated electronic systems. In this paper we present the results of these investigations.

In the next section~\ref{sec:2} we consider the Hubbard model of an electronic system  with strong  correlations and explain how the kinematical interaction  appears  in the  model. Then, we  describe the theory of  superconductivity developed   in our studies.  To disclose the mechanism of HTSC, we
consider in section~\ref{sec:3} the extended Hubbard model where we can compare contributions from the spin-fluctuations and the electron-phonon interaction (EPI) and evaluate a role of the intersite Coulomb interaction (CI). The exact Dyson equation for the normal and anomalous (pair) GFs is derived for this model which is solved in the self-consistent Born approximation (SCBA) for the self-energy. In   section  \ref{sec:4} the equation for the superconducting order parameter  is considered. We found the $d$-wave pairing with
high-$T_c$ mediated by the kinematical electron interaction with spin-excitations. $T_c$
can be suppressed only for a large intersite CI of the order of kinematical interaction,
much  larger than EPI. This supports the spin-fluctuation mechanism of high-$T_c$ in
cuprates. In conclusion we summarize our results.

\section{Kinematical interaction }
\label{sec:2}

The problem of the many-body effects caused by strong electron correlations in solids  is
one of the most difficult ones which has not yet found   a comprehensive solution. Strong
electron correlations are usually treated within specific models in which only few
relevant electronic states are taken into account (for a review see~\cite{Fulde95,Avella11}).
In the simplest approximation, one can consider a one-band model and take into account
only a single-electron hopping matrix element $t$ between the nearest neighbors and a
large single-site Coulomb energy $U$. In this approximation, the complicated  multiband electronic  model of cuprates is reduced to the so-called Hubbard model~\cite{Hubbard63}:
\begin{equation}
H= -t \sum_{i \neq j\sigma}\,  a_{i\sigma}^{\dag}\, a_{j\sigma} + U \, \sum_{i}
N_{i\uparrow} \, N_{i\downarrow},
 \label{1}
\end{equation}
where $a^{\dag}_{i\sigma}$($a_{i\sigma}$) are the  creation (annihilation) operators for
electrons of spin $\sigma = (\uparrow, \downarrow) \equiv (+, -)\,$ at the lattice site
$i$  and $N_{i\sigma} = a^{\dag}_{i\sigma} a_{i\sigma}$ is the electron occupation
number. In the limit of strong correlation, $U \gg t$,  the conduction band  splits into
two subbands for the singly occupied and doubly occupied states.  In
this case, to describe the electronic excitations, one has to use the projected electron
operators referring to these subbands: $\, a_{i\sigma} =  a_{i\sigma}(1- N_{i
\bar\sigma})  + a_{i\sigma} N_{i \bar\sigma} \equiv  \widetilde{a}_{i\sigma} +
\widetilde{d}_{i\sigma},  \,$ where $ \bar\sigma = - \sigma$.  It is convenient to use
the HO representation~\cite{Hubbard65}:
\begin{equation}
\widetilde{a}_{i\sigma} =  a_{i\sigma}(1- N_{i \bar\sigma}) \rightarrow  X_{i}^{0 \sigma
}, \quad \widetilde{d}_{i\sigma}= a_{i\sigma} N_{i \bar\sigma} =
a_{i \bar\sigma}^{\dag} a_{i \bar\sigma}\,a_{i\sigma} \rightarrow X_{i}^{\bar\sigma 2} .
 \label{2}
\end{equation}
The matrix of HOs $X_{i}^{\alpha\beta} = |i\alpha\rangle\langle i\beta|$ describes the
transition from the state $|i,\beta\rangle$ to the state $|i,\alpha\rangle$ on a lattice
site $i$ taking into account four  possible electronic states: an empty state $(\alpha,
\beta =0) $, a singly occupied electronic state $(\alpha, \beta = \sigma)$, and a doubly
occupied electronic state $(\alpha, \beta = 2) $. The number operator and the spin
operators in terms of the HOs are defined as
\begin{eqnarray}
  N_i &=& \sum_{\sigma} X_{i}^{\sigma \sigma} + 2 X_{i}^{22},
\label{3}\\
S_{i}^{\sigma} & = & X_{i}^{\sigma\bar\sigma} ,\quad
 S_{i}^{z} =  (\sigma/2) \,[ X_{i}^{\sigma \sigma}  -
  X_{i}^{\bar\sigma \bar\sigma}] .
\label{4}
\end{eqnarray}
From the multiplication rule for the HOs, $\, X_{i}^{\alpha\beta} X_{i}^{\gamma\delta} =
\delta_{\beta\gamma} X_{i}^{\alpha\delta}$, there follow their commutation relations
\begin{equation}
\left[X_{i}^{\alpha\beta}, X_{j}^{\gamma\delta}\right]_{\pm}=
\delta_{ij}\left(\delta_{\beta\gamma}X_{i}^{\alpha\delta}\pm
\delta_{\delta\alpha}X_{i}^{\gamma\beta}\right)\, ,
 \label{5}
\end{equation}
with the upper  sign  for the Fermi-type operators (such as $X_{i}^{0\sigma}$) and the
lower sign for the Bose-type operators (such as  the number  or  spin operators). The HOs
obey the completeness relation
\begin{equation}
 X_{i}^{00} +
 \sum_{\sigma} X_{i}^{\sigma\sigma}  + X_{i}^{22} = 1,
 \label{5a}
\end{equation}
which rigorously preserves the local constraint  that only one quantum state $\alpha$ can
be occupied on any lattice site $i$. This no-double-occupancy   restriction is the most
difficult property of the projected electronic operators which is considered in many
theoretical approaches in mean-field approximation (MFA) resulting in unjustified
results.

The unconventional commutation relations (\ref{5}) for HOs result in the {\bf kinematical
interaction}. The term was introduced by Dyson in a general theory of spin-wave
interactions~\cite{Dyson56} for the spin-wave creation $\,
b_i^{\dag}= S_{i}^{-} $ and annihilation   $\, b_i =S_{i}^{+} \,$ operators for
spin-$1/2$.  It appears that they are Bose operators on different lattice sites and Fermi operators on the same lattice site. Similar commutation relations  hold for the electron creation
$X_{i}^{\sigma 0}\, $ and annihilation $ X_{j}^{0\sigma}\,$ operators,
\begin{align}
 X_{i}^{0\sigma} X_{j}^{\sigma 0}+ X_{j}^{\sigma 0}X_{i}^{0\sigma}
 = \delta_{i,j}(1 - X_{i}^{\bar\sigma \bar\sigma} - X_i^{22})
% \nonumber \\
  = \delta_{ij}(1 - N_{i \sigma} /2 + \sigma S_i^z).
 \label{7}
\end{align}
Therefore, the HOs for electrons  can be considered as Fermi operators on different lattice sites but on the same lattice site electron kinematical interaction  occurs with charge  $N_{i \sigma}$ and spin $S_i^\alpha$ fluctuations. This results in dressing  the electron hopping by spin and charge fluctuations with the  coupling constant of the order of the kinetic energy $\, W \sim 4 t$ (for a two
dimensional lattice). The latter is much larger than the antiferromagnetic exchange interaction $J \sim 4t^2/U$  proposed by Anderson~\cite{Anderson87}  as the coupling parameter for superconducting pairing in cuprates.  The superconducting pairing induced by the kinematical interaction for the HOs was first proposed  by Zaitsev and Ivanov~\cite{Zaitsev87, Zaitsev87a,Zaitsev87b}. However, they found in the MFA the momentum-independent $s$-wave superconducting gap  which  violates the no-double-occupancy   restriction  as was shown in references~\cite{Plakida89,Yushankhai91}. The spin-fluctuation pairing in the second order of the kinematical interaction  beyond the MFA should be taken into account which results in the $d$-wave pairing as discussed below.

\section{Green function equations}
\label{sec:3}

We apply our theory for consideration of HTSC in cuprate superconductors. To compare
electron-phonon and spin-fluctuation pairing mechanisms, we consider the extended  Hubbard
model which includes the intersite CI $V_{ij}$ and the EPI $g_{ij}$. In terms of the HOs,
the model reads:
\begin{eqnarray}
 H &= & \varepsilon_1\sum_{i,\sigma}X_{i}^{\sigma \sigma}
  + \varepsilon_2\sum_{i}X_{i}^{22} + \sum_{i\neq j,\sigma}\,
t_{ij}\,\bigl\{ X_{i}^{\sigma 0} \, X_{j}^{0\sigma}
\nonumber \\
& + &    X_{i}^{2 \sigma}X_{j}^{\sigma 2}
 + \sigma \,(X_{i}^{2\bar\sigma} X_{j}^{0 \sigma} +
  {\rm H.c.})\bigr\} + H_{c, \text{ep}}\,,
 \label{8}
\\
H_{c, \text{ep}} &= & \frac{1}{2}  \sum_{i\neq j}\,V_{ij} N_i N_j + \sum_{i, j}\,g_{i j} N_i\,
u_j,
 \label{8a}
\end{eqnarray}
where $t_{i,j}$ is the single-electron hopping parameter between lattice sites $i$  and
$j$, $\varepsilon_1 = - \mu$ is the single-particle energy  and $\varepsilon_2 =  U - 2
\mu $ is the two-particle energy, $\mu$ is the chemical potential. $u_j$ describes  an
atomic displacement on the lattice site $j$ for a particular phonon mode. We emphasize
that in the Hubbard model (\ref{8}) there is no dynamical coupling of electrons
(holes) with spin or charge fluctuations. Its role is played by the  kinematical
interaction  as was discussed in the previous section.

To consider the superconducting pairing in the model (\ref{8}), we introduce the
thermodynamic GF~\cite{Zubarev60} using  the four-component Nambu operators, $\, \hat
X_{i\sigma}$ and   $\, \hat X_{i\sigma}^{\dagger}=(X_{i}^{2\sigma}\,\, X_{i}^{\bar\sigma
0}\,\, X_{i}^{\bar\sigma 2}\,\, X_{i}^{0\sigma}) \,$:
\begin{eqnarray}
 {\sf G}_{ij\sigma}(t-t') = -\ri \theta(t-t')\langle \{
 \hat X_{i\sigma}(t) ,  \hat X_{j\sigma}^{\dagger}(t')\}\rangle
 \equiv \langle \!\langle \hat X_{i\sigma}(t) \mid
    \hat X_{j\sigma}^{\dagger}(t')\rangle \!\rangle,
 \label{9}
\end{eqnarray}
where $ \{A, B\} = AB + BA$,  $ A(t)= \exp (\ri Ht) A\exp (-\ri Ht)$, $\langle AB\rangle$ is
the statistical average of operators $AB$, and $\theta(x) $ is the Heaviside function.
The Fourier representation of the GF (\ref{9}) in the  $({\bf k}, \omega) $-space is defined by the relation:
\begin{eqnarray}
{\sf G}_{ij\sigma}(t-t') =\int_{-\infty}^{\infty} \frac{\rd t}{2\piup} \re^{- \ri\omega(t-t')}
\frac{1}{N}\,\sum_{\bf k}\exp[\ri{\bf k (i-j)}] {\sf
G}_{\sigma}({\bf k}, \omega).
    \label{10}
\end{eqnarray}
The Fourier components of the GF (\ref{10}) are convenient to write in the matrix form
\begin{equation}
{\sf G}_{\sigma}({\bf k}, \omega)=
  {\hat G_{\sigma}({\bf k}, \omega)  \quad \quad
 \hat F_{\sigma}({\bf k}, \omega) \choose
 \hat F_{\sigma}^{\dagger}({\bf k}, \omega) \quad
   -\hat{G}_{\bar\sigma}(-{\bf k}, -\omega)} ,
 \label{11}
\end{equation}
where the normal $\hat G_{\sigma}({\bf k}, \omega)$ and anomalous (pair) $\hat
F_{\sigma}({\bf k}, \omega) $  GFs are  $2\times 2$ matrices for the two Hubbard subbands.

To calculate the GF (\ref{9})  we use the projection technique in the equation of motion method~\cite{Plakida11} by differentiating the GF  with respect to  time $t$ and $t'$, similar to the Mori projection technique~\cite{Mori65}. This method  can be applied  to any type of operators since it is not based on any diagram technique. A  general theory of superconductivity within this method is presented in \cite{Plakida10a}.  As a result, we derive the Dyson equation in the form (for details see~\cite{Plakida13,Plakida16}):
\begin{equation}
 {\sf G}\sb{\sigma}({\bf k}, \omega) =
  \left[\omega \tilde{\tau}_{0} - {\sf E}_{\sigma}({\bf k})
  -    {\sf Q}\,\Sigma_{\sigma}({\bf k}, \omega)
  \right] \sp{-1} {\sf Q},
\label{12}
\end{equation}
where $\tilde{\tau}_{0}$ is the $4 \times 4$ unit matrix. ${\sf E}\sb{\sigma}({\bf k})$
is  the quasiparticle (QP) electronic excitation energy in the generalized MFA  given by
the matrix:
\begin{equation}
 {\sf E}\sb{\sigma}({\bf k})=
 ({1}/{N})\sum_{\bf i , j}\exp[\ri{\bf k (i-j)}] \langle \{ [\hat X\sb{i\sigma}, H],
\hat X\sb{j\sigma}\sp{\dagger} \} \rangle {\sf Q}^{-1}.
 \label{12a}
\end{equation}
The  correlation function ${\sf Q} = \langle \{\hat X\sb{i\sigma},\hat
X\sb{i\sigma}\sp{\dagger}\}\rangle   = \hat{\tau}_{0} \times \hat Q\,$ determines the
spectral weights of the Hubbard subbands  where   $\, \hat
Q = \left(  \begin{array}{cc} Q\sb{2} & 0 \\
0 & Q\sb{1} \end{array}  \right)\,$ and $\hat{\tau}_{0}$ is the $2 \times 2$ unit matrix.
In the paramagnetic state  $\, Q\sb{2} = n/2 \,$ and $\, Q\sb{1} = 1- n/2\, $ depend on
the average occupation number of electrons $n = \langle N_i\rangle$.

The self-energy operator is given by the multiparticle GF
\begin{equation}
 {\sf Q}\,\Sigma_{\sigma}({\bf k}, \omega)=
    \langle\!\langle {\hat Z}\sb{{\bf k}\sigma}\sp{(\rm ir)} \!\mid\!
     {\hat Z}\sb{{\bf k}\sigma}\sp{(\rm ir)\dagger} \rangle\!\rangle
      \sp{(\rm pp)}\sb{\omega}\;{\sf  Q}\sp{-1} ,
\label{13}
\end{equation}
where $\hat Z\sb{i\sigma}\sp{(\rm ir)} = [\hat X\sb{i\sigma}, H] - \sum\sb{l}{\sf
E}\sb{il\sigma} \hat X\sb{l\sigma} $ and the proper part $(\rm pp)$ notation means that the multiparticle GF cannot be cut into two parts connected by the singleparticle GF.  The self-energy operator (\ref{13}) can be written in the same matrix form as the  GF (\ref{11}):
\begin{equation}
 {\sf Q}\,\Sigma_{\sigma}({\bf k}, \omega) =  {\hat M_{\sigma}({\bf k}, \omega)
 \quad  \quad
\hat\Phi_{\sigma}({\bf k}, \omega) \choose \hat\Phi_{\sigma}^{\dagger} ({\bf k},
\omega)\quad -\hat{M}_{\bar\sigma}({\bf k}, -\omega)} {\sf Q}^{-1}  \, ,
 \label{14}
\end{equation}
where the matrices $\hat M$ and $\hat\Phi$  denote the corresponding normal and anomalous
(pair) components of the self-energy operator.
As a result, we obtain an exact representation for the GF (\ref{9}) in the Dyson equation~(\ref{12})  detemined by the zero-order QP excitation energy (\ref{12a}) and the
self-energy (\ref{14}).  The self-energy takes into account processes of inelastic scattering of electrons (holes) on spin  and charge fluctuations due to the kinematical interaction and the dynamic
intersite CI and the EPI.

We calculate the self-energy  (\ref{13}) in the SCBA. Let us consider, for instance, the anomalous component of the self-energy  for the the doubly occupied  subband:
\begin{equation}
 \Phi^{22}_{l l',\sigma}( \omega) = \langle\langle [X_l^{\sigma 2}, H ] |[H, X_{l'}^{\bar{\sigma} 2}] \rangle\rangle_{\omega} \, .
 \label{15}
\end{equation}
Performing commutations of the HOs with the Hamiltonian (\ref{8}) in  $ [X_l^{\sigma 2}, H ]$,
we obtain for the self-energy (\ref{15})  such terms as
$ \,\langle\langle X_l^{\sigma' 2}B\sb{i\sigma \sigma'} | X_{l'}^{\bar{\sigma}' 2} B\sb{j
\bar{\sigma}\bar{\sigma}'}\rangle\rangle_{\omega} \, $ where the Bose-type operator $ B\sb{i\sigma\sigma'} = ( N\sb{i}/2 +  \sigma\, S\sb{i}\sp{z}) \, \delta\sb{\sigma'\sigma} +
S\sb{i}\sp{\sigma} \, \delta\sb{\sigma'\bar\sigma}$ is induced by the kinematical interaction  in equation (\ref{7}). According to the spectral representation for the GFs~\cite{Zubarev60}, they  can be written in terms of the corresponding two-time correlation function, in our case, $ \langle  X_{l'}^{\bar{\sigma}' 2} B\sb{j \bar{\sigma}\bar{\sigma}'}  | B\sb{i\sigma \sigma'}(t) X_l^{\sigma' 2}(t)\rangle$.  In the SCBA, a propagation of excitations described by the Fermi-type operators $\,X_l^{\sigma' 2}(t) \,$ and  the Bose-type operators
$ B\sb{i\sigma\sigma'}(t)$ on different lattice sites $l \neq i$ is assumed to be
independent. Therefore, the multiparticle correlation function can be written   as a product
of fermionic and bosonic time-dependent correlation functions:
\begin{eqnarray}
\langle  X_{l'}^{\bar{\sigma}' 2} B\sb{j
\bar{\sigma}\bar{\sigma}'}
 | B\sb{i\sigma \sigma'}(t) X_l^{\sigma' 2}(t)\rangle
 =  \langle  X_{l'}^{\bar{\sigma}' 2}
  X_l^{\sigma' 2}(t)\rangle\,
  \langle   B\sb{j\bar{\sigma}\bar{\sigma}'}
  B\sb{i\sigma \sigma'}(t) \rangle \, .
 \label{15a}
\end{eqnarray}
The fermionic correlation function $ \langle  X_{l'}^{\bar{\sigma}' 2}   X_l^{\sigma' 2}(t)\rangle $ is  self-consistently calculated from the corresponding GF  $ \langle \langle X_l^{\sigma' 2} |  X_{l'}^{\bar{\sigma}' 2} \rangle \rangle_{\omega} $. The bosonic correlation function $ \langle   B\sb{j\bar{\sigma}\bar{\sigma}'}   B\sb{i\sigma \sigma'}(t) \rangle $ can be written in terms of the dynamical spin susceptibility $\chi_{\text{sf}}({\bf q}, \omega)= -  \langle\langle {\bf S_{\bf q}} | {\bf S_{-\bf q}} \rangle\rangle_{\omega}\,$ and the dynamical charge susceptibility $\chi_{\text{ch}}({\bf q}, \omega)= -  \langle\langle { N_{\bf q}} | { N_{-\bf q}} \rangle\rangle_{\omega}\,$.  The same SCBA is used to calculate the normal components of the self-energy $\,\hat M_{\sigma}({\bf k}, \omega)\,$. In this approximation, we obtain the self-consistent system of equations for the GF  (\ref{12}) and the the self-energy (\ref{14}).

\section{Superconductivity in  the extended  Hubbard model}
\label{sec:4}

The electronic spectrum in the normal state is determined by the normal components of the  matrix~(\ref{12a}) and  the self-energy matrix $\hat M_{\sigma}({\bf k}, \omega)$ in equation~(\ref{14}). The spectrum was considered in detail in references~\cite{Plakida13,Plakida14,Plakida16}.  Therefore, here we
discuss only the superconductivity in the model. Let us consider  the equation for the
superconducting order parameter, the gap function, e.g., in the doubly occupied hole
subband. The gap is determined by the anomalous contribution in equation~(\ref{12a}),
$\Delta^{22}_{ij\sigma}  = (J_{i j} - V_{i j})\,\langle X_{i}^{\sigma2}
 X_{j}^{\bar\sigma2}\rangle /Q_2 ,$  and the anomalous  self-energy component $\, \Phi^{22}_{\sigma}({\bf k}, \omega)\,$ (\ref{15}):
 \begin{equation}
  \varphi_\sigma({\bf k},\omega)=  \Delta^{22}_{\sigma}({\bf k}) +
 \Phi^{22}_{\sigma}({\bf k},\omega) /Q_2 .
  \label{16}
\end{equation}
 To calculate the superconducting $T_c$, we consider a simplified version of the linearized equation for the  gap function (\ref{16}) at the Fermi energy $\varphi({\bf k}) = \sigma \, \varphi_\sigma({\bf
k},\omega=0)$:
\begin{eqnarray}
\varphi({\bf k}) &=&
  \frac{1}{N}\sum_{{\bf q}} \,
  \frac{  \varphi({\bf q})}{[Z({\bf q})]^2 \; 2\widetilde{\varepsilon}({\bf q})}
    \tanh\frac{\widetilde{\varepsilon}({\bf q})}{2T_c}
 \big\{ J({\bf k-q}) - V({\bf k-q})
\nonumber \\
&+& \big[(1/4) |t({\bf q})|\sp{2} + |V({\bf k -q})|^2\big] \chi\sb{\text{cf}}({\bf k -q})
 \nonumber \\
& +& |g({\bf k -q})|^2 \, \chi_{\text{ph}}({\bf k-q})\,
 -  |t({\bf q})|\sp{2}\; \chi_{\text{sf}}({\bf k -q}) \big\}\, ,
     \label{17}
\end{eqnarray}
where in the static approximation for the self-energy we introduce  the static
charge $\chi\sb{\text{cf}}({\bf k -q})$, spin $\chi\sb{\text{sf}}({\bf k -q})$,
and phonon  $\chi_{\text{ph}}({\bf k-q})$ susceptibilities.
The renormalized electronic energy $\,\widetilde{\varepsilon}({\bf q}) = \varepsilon_{2}({\bf q})/Z({\bf q}) $ is determined by
the renormalization parameter $Z({\bf q}) = 1 - [\partial M^{22}({\bf q},\omega)/\partial
\omega]_{\omega =0}$ related to  the QP weight $1/Z({\bf q})$ in the normal state.

To estimate various contributions in the gap equation (\ref{17}), we consider a model
$d$-wave gap function $\varphi({\bf k}) = (\Delta/2) \,\eta({\bf k})$ where $ \eta({\bf
k}) =(\cos k_x - \cos k_y)$. Integrating equation~(\ref{17}) with the function $\eta({\bf
k})$ over $ {\bf k}$ we obtain the  gap equation  in the form:
\begin{eqnarray}
 1 &=& \frac{1}{N } \sum_{\bf q}\frac{  [\eta({\bf q})]^2}{[Z({\bf q})]^2 \;
  2\widetilde{\varepsilon}({\bf q})}
  \tanh\frac{\widetilde{\varepsilon}({\bf q})}{2T_c}
  \big\{ J -\widehat{V}_{\text{c}} + \widehat{V}_{\text{ep}}
 \nonumber \\
 & +& \widehat{V}\sb{\text{cf}}  +(1/4)\,|t({\bf q})|\sp{2} \widehat{\chi}\sb{\text{cf}}
-  |t({\bf q})|^{2}\, \widehat{\chi}\sb{\text{sf}}
    \big\}.
\label{18}
\end{eqnarray}
Here,  we take into account that $ \sum_{\bf k}\eta ({\bf k}) F({\bf k- q}) = \sum_{\bf k}F({\bf k}) \,\eta ({\bf k + q}) = \sum_{\bf k} F({\bf k}) [\cos (k_x +  q_x) -   \cos (k_y +q_y) ] = \eta ({\bf q}) \sum_{\bf k} F({\bf k})\cos k_x \, $ for functions in the tetragonal phase, $F({k_x, k_y}) = F({k_y, k_x})$. Therefore, for the $d$-wave pairing only $l = 2 $ symmetry components  of the interaction functions $ F({\bf k}) $ give contributions:
\begin{eqnarray}
 \widehat{V}_{\text{c}} & = & \frac{1}{N}\sum_{\bf k} V({\bf k})\cos k_x,
\quad \widehat{V}_{\text{ep}}=  \frac{1}{N}\sum_{\bf k}
 g({\bf k})\cos k_x,
\nonumber \\
\widehat{\chi}\sb{\text{cf}} & = &\frac{1}{N}\sum_{\bf k}\chi\sb{\text{cf}}({\bf k}) \cos k_x , \quad
\widehat{V}\sb{\text{cf}} =\frac{1}{N}\sum_{\bf k}
   |V({\bf k})|^2\chi\sb{\text{cf}}({\bf k})\,\cos k_x,
\nonumber \\
    \widehat{\chi}_{\text{sf}} & = &  \sum_{{\bf k}} \chi_{\text{sf}}({\bf k})\, \cos k_x =
   \sum_{{\bf k}} \frac{\chi_{\bf Q} }{ 1+\xi^2[1+\gamma({\bf k})]} \cos k_x  <  0 .
  \label{19}
\end{eqnarray}
For the intersite CI we consider the model  $\,V({\bf k}) = 2 V\, (\cos k_x + \cos k_y ) \,$ with various values of  $\, 0 \leqslant V \leqslant 2\,t\,$ (see reference~\cite{Plakida14}).   EPI is described by the model
$\,g ({\bf k})= g_{\text{ep}}/(\kappa^2 + |{\bf k}|^2)\,$ where we take a large coupling constant $\, g_{\text{ep}} = 5\,t$. The screening constant depends on  doping, $\, \kappa =  2 \delta\,$,  and determines  the strong EPI at a small doping, while it is suppressed at a large doping  (see reference~\cite{Plakida13}).

For the AF spin susceptibility  $ \chi_{\text{sf}}({\bf k})$ in  (\ref{19}) we take a model function  with a
peak   at the AF wave vector ${\bf  Q} = (\piup,\piup)$.  The strength of the spin-fluctuation interaction
is  determined  by the static susceptibility $\chi_{\text{sf}}({\bf Q}) =\chi_{\bf Q} $ for $\,[ 1+\gamma({\bf Q})] = 0\,$. It is calculated from the normalization condition $\langle {\bf S}_{i}^2\rangle
  = ({3}/{4})(1- \delta)$ and is given by the equation (see reference~\cite{Plakida13}):
\begin{equation}
\chi_{\bf Q}  = \langle {\bf S}_{i}^2\rangle\frac{2}{\omega_{s} }
 \left\{ \frac{1}{N} \sum_{\bf q}
 \frac{1}{ 1+\xi^2[1+\gamma({\bf q})]} \right\}^{-1} .
 \label{52}
 \end{equation}
The model is determined by two parameters: the AF correlation length $\xi$ which depends on  doping and  the characteristic energy of spin excitations of the order of the exchange energy $\omega_s \sim J$. For the doping dependence of the AF correlation length (in units of the lattice parameter $a$), we use a simple approximation observed in neutron scattering   experiments $\xi \approx  1/\sqrt{\delta}$. For a small doping, $\xi$ is large and the spin-fluctuation interaction  $\chi_{\bf Q}$ is strong, while for a large doping, the parameter  $\chi_{\bf Q}$ decreases and the spin-fluctuation interaction becomes weak. This dependence  partially explains the variation of $T_c$ induced by  spin-fluctuations. The doping dependence of  the density of electronic states  also influences the $T_c$ dependence induced   both by the spin fluctuation interaction and EPI. Note that the negative sign of the spin-fluctuation contribution in equations~(\ref{17}) and (\ref{18}) is compensated by the  negative sign of the parameter $\widehat{\chi}_{\text{sf}}$ since the main contribution in equation~(\ref{19}) comes  at the AF wave vector ${\bf Q}$ where $\cos k_x  < 0$.
\begin{figure}[!t]
\centerline{\includegraphics[width=0.4\textwidth]{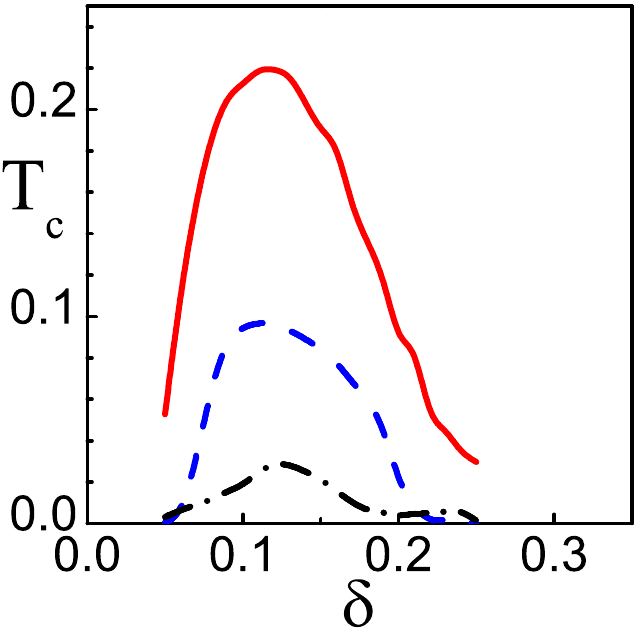}}
\caption{(Color online)  $T_c(\delta)$ in units of $t$ in the WCA induced by all
interactions (red solid line) and only by the spin-fluctuation contribution
$\widehat{\chi}\sb{\text{sf}}$ (blue dashed line) or only by the EPI $\, \widehat{V}\sb{\text{ep}}$
(black dash-dotted line). }
 \label{fig1}
\end{figure}

To compare  various mechanisms of pairing, let us consider the solution of equation~(\ref{18}) for $T_c$ as a function of hole doping $\delta$ in the weak coupling approximation (WCA) neglecting the self-energy contribution  by taking $Z({\bf q}) = 1$.  In WCA we obtain high values for $T_c$  as shown in figure~\ref{fig1}.
A very high $T_c \approx 0.22\, t $ is found when all the contributions are taken into
account. The spin-fluctuation pairing results in superconducting $T^{\text{sf}}_c \approx 0.1\,
t  $ which is  much larger than $T^{\text{ep}}_c \approx 0.02 t$ mediated only by  EPI. The
doping dependence of $T_c$  qualitatively agrees with experiments in cuprates but its
value is too high. It is important to stress that contributions from CI $\widehat{V}_{\text{c}}$ and EPI $\widehat{V}_{\text{ep}}$ are suppressed since they are given by $l = 2$  component of the interactions.  In particular, EPI may be strong for  phonon modes with ${\bf k }$-independent interaction resulting in $\widehat{V}_{\text{ep}} = 0$ but  providing a pronounced polaronic effect observed in experiments. Similar contribution from single-site CI $U$ is zero. We also note that the pairing in the MFA induced by the AF exchange $J$ proposed by Anderson~\cite{Anderson87} provides quite a low $T_c$ which is suppressed by the intersite CI since $\, J - \widehat{V}_{\text{c}} \approx 0.1 t$.
\begin{figure}[!t]
\centerline{\includegraphics[width=0.4\textwidth]{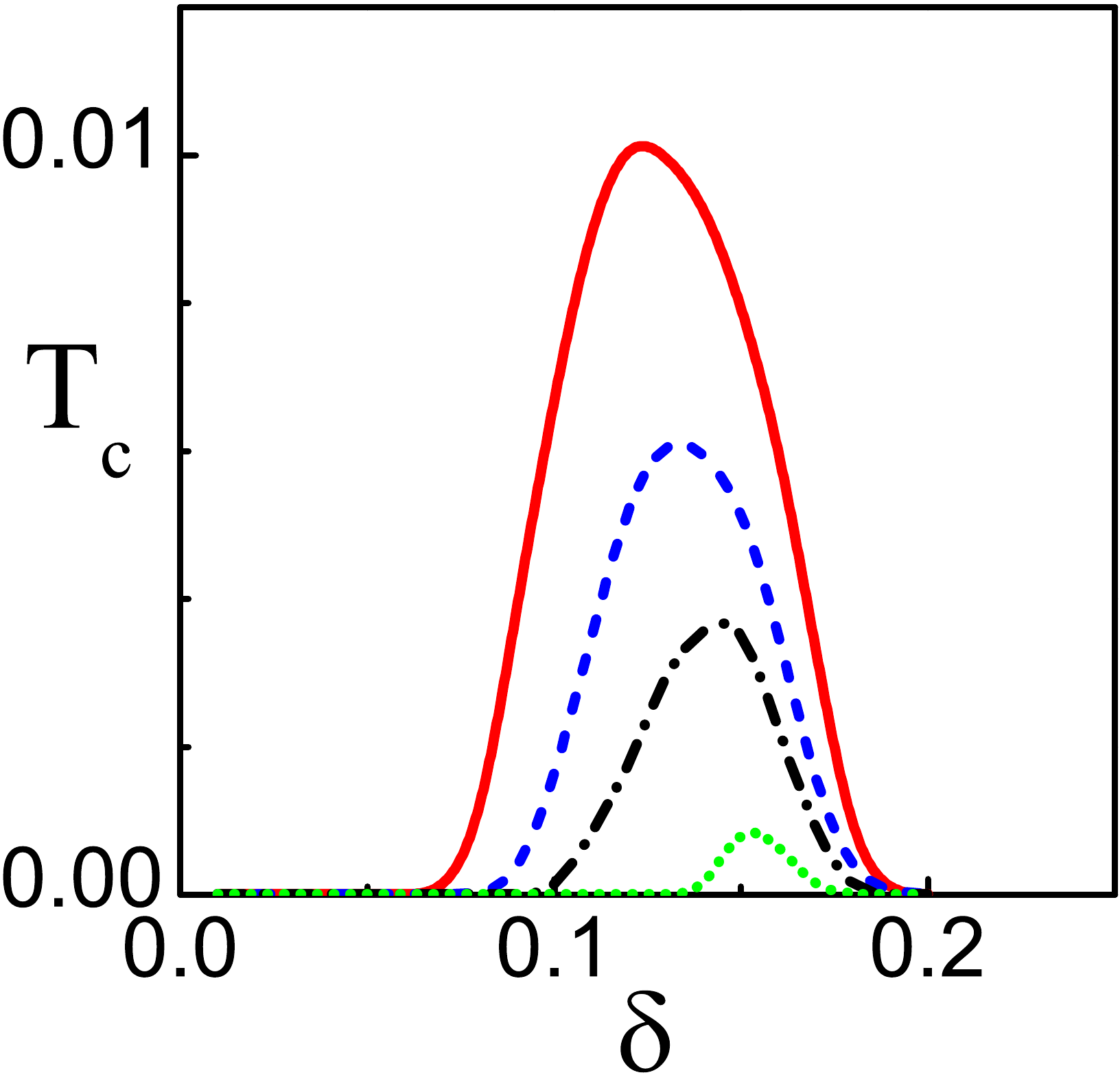}}
\caption{(Color online)  Solution of equation~(\ref{17}) for
 $T^{\text{sf}}_c(\delta)$  for the CI $\widehat{V}_c =
0.0$ (red solid line), $\widehat{V}_c = 0.5 $ (blue dashed line), $\widehat{V}_c = 1.0 $
(black dash-dotted line), and $\widehat{V}_c = 2.0 $ green dotted line).}
 \label{fig2}
\end{figure}

Now, we consider the solution of equation~(\ref{17}) taking into account the  renormalization parameter $Z({\bf q})$  which is quite large, $Z({\bf q}) = 3-4 $,  as shown in references~\cite{Plakida13,Plakida14}. In this case we obtain the values of  $T_c$ that are an order of magnitude smaller. Figure~\ref{fig2} shows the solution of equation~(\ref{17}) for $T^{\text{sf}}_c$ in the presence of the intersite CI. $T^{\text{sf}}_c$ for $\widehat{V}_{\text{c}} = 0.0$ is quite high due  to a strong coupling induced by the kinematical interaction $|t({\bf q})|\sp{2}$ and is close to experimentally observed one, $T^{\text{sf}}_c \sim 100$~K,  for the characteristic value of $t = 0.4$~eV. Increasing  $\widehat{V}_{\text{c}}$ suppresses $T_c$ which becomes small only for high values of $\widehat{V}_{\text{c}} = 2t - 3t$ comparable with the spin-fluctuation coupling $\, g_{sf} \sim 4t$ and is much larger than the EPI and the AF exchange interaction $\, J \sim 0.4 t$.

\begin{figure}[!t]
\centerline{\includegraphics[width=0.4\textwidth]{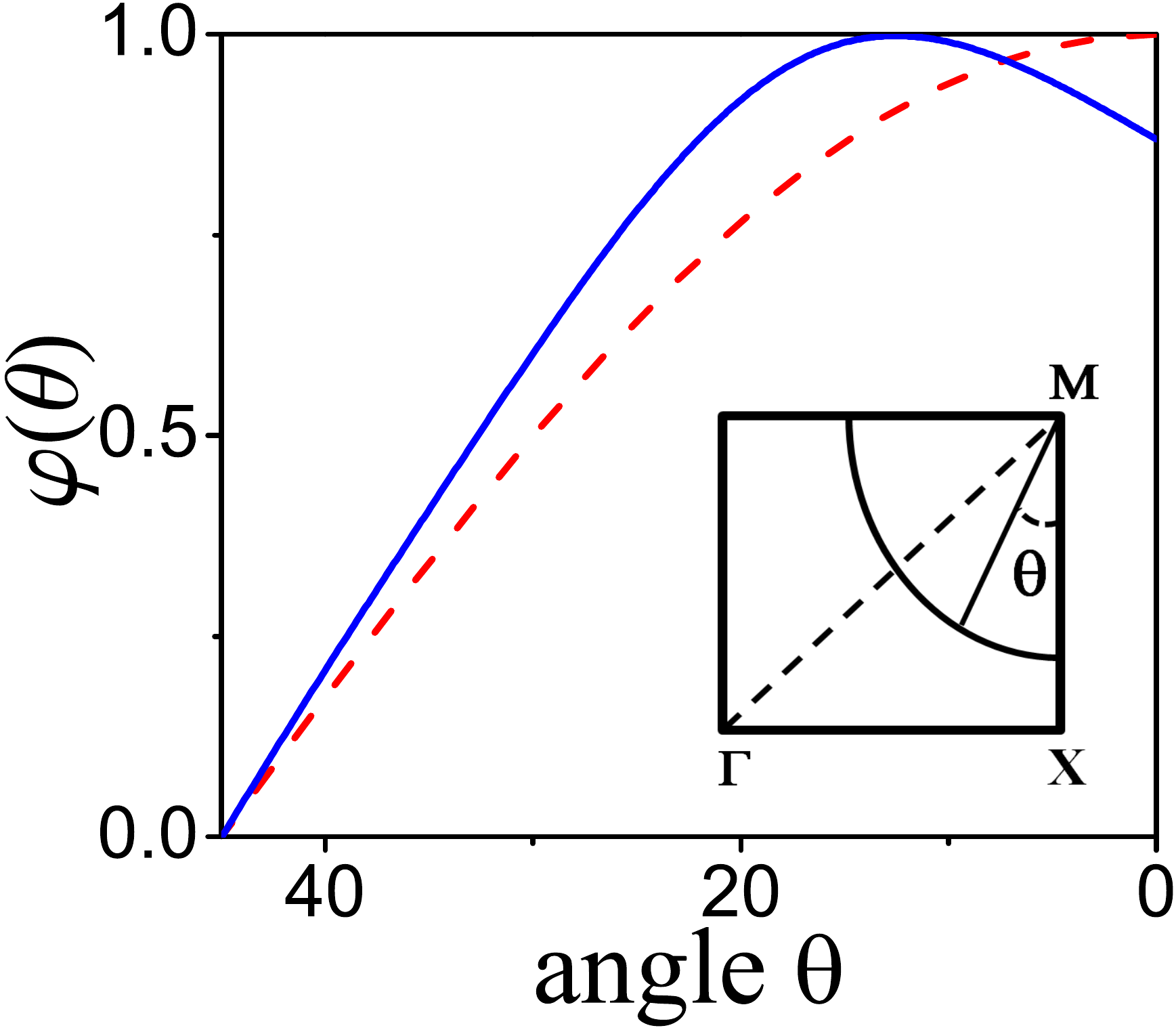}}
 \caption{(Color online) Angle dependence of the SC gap $\varphi(\theta) $ on the FS (blue bold line) in comparison with the model $d$-wave dependence $\varphi_d(\theta) = \cos 2\theta$ (red
dashed lines).}
 \label{fig3}
\end{figure}

As discussed in~\cite{Plakida13,Plakida14,Plakida16}, numerical solution of the
full gap equation shows the energy gap dependence   characteristic of the pairing induced by
bosons, in our theory --- spin-fluctuations.  The ${\bf k}$-dependence of the gap function  $\varphi({\bf k})\,$ on the Fermi surface at doping $\delta = 0.13$ shown in figure~\ref{fig3} reveals the  $d$-wave symmetry which  angle dependence is close to the model $d$-wave dependence $\varphi_d(\theta) = \cos 2\theta$.
By taking into account the electron-phonon interaction within the extended
Hubbard model (\ref{8})  the isotope effect (IE) was found~\cite{Plakida16}. The IE exponent  $\alpha =  -\rd \ln T_{\rm c}/ \rd \ln M $ is small, $\alpha \leqslant 0.1 $,  at optimal doping, while  in the underdoped region, the IE increases and can be even larger than in the conventional superconductors, $\alpha \sim 1 \, $,  which is  in a qualitative agreement with experiments.

\section{Conclusion}
\label{sec:5}

The microscopic theory of superconductivity of strongly correlated electronic systems such as cuprates is formulated.  The superconducting pairing  is provided by strong kinematical interaction of  electrons with spin excitations. The EPI and the intersite CI are suppressed for the $d$-wave pairing since only $l = 2 $ symmetry components  of the interactions give contributions.  We emphasize that there is no kinematical interaction in the phenomenological  spin-fermion models where the conventional
electrons interacting with spin fluctuations are considered.  The kinematical interaction
is lost in the slave-boson (-fermion) models treated in the MFA where the projected
electron operators are approximated by a product of  conventional electron and  boson operators
in MFA: $\widetilde{a}_{i\sigma}  = a_{i\sigma}  b_i = a_{i\sigma} \langle b_i \rangle$.

The spin-fluctuation mechanism was supported in many publications, in particular, for systems with strong electron correlations. Here, we refer to numerical simulations for finite clusters (see
reviews~\cite{Scalapino95,Bulut02,Scalapino07,Senechal12,Scalapino12}),  the dynamical cluster approximation
(DCA)~\cite{Maier05,Maier06,Maier06a}, and the cluster dynamical mean-field theory (see, e.g.,
\cite{Capone06,Haule07,Kancharla08}).  Extensive numerical studies for finite clusters have revealed a tendency to the $d$-wave pairing in the Hubbard model at large $U \gg t$, though a delicate balance between superconductivity and other instabilities (AF, spin-density wave, charge-density wave,
etc.) was found in the above cited references. In~\cite{Macridin09},  using the DCA with the quantum Monte Carlo method, the
superconducting $d$-wave pairing and the isotope effect similar to the one observed in cuprates
were found for the Hubbard-Holstein model. However, in some finite cluster calculations, an appearance
of the long-range superconducting order has not been confirmed (see, e.g., \cite{Aimi07}). The drawback of this result  may be due to a finite number of electrons in clusters.

Our  conclusion concerning the importance of the kinematical mechanism of pairing is
supported by the studies in~\cite{Plekhanov03}. Using the variational Monte Carlo technique the superconducting $d$-wave  gap was observed for the extended Hubbard model with a weak exchange interaction $J = 0.2 \, t$  and a repulsion $V \leqslant 3\, t $ in a broad range of $\,0 \leqslant U \leqslant 32$. It was found that the gap decreases with increasing $V$ at all $U$ and can be suppressed for  $V > J $ for small $\, U $. However, for large $ \,U \gtrsim U_c \sim 6\, t\,$ the gap becomes  robust and exists up to  large values of $V \sim 10\, J = 2\, t $.  We can give an  explanation of these results by pointing out that at large $U \gtrsim U_c $, the concomitant decrease of the bandwidth  results in the splitting of the Hubbard band into the upper and lower subbands, and the emerging kinematical interaction induces the $d$-wave pairing in one Hubbard subband. In that case, the second subband for large $U$
gives a small contribution which results in the $U$-independent pairing.  This can be suppressed by the repulsion $V$ being only larger than the kinematical interaction, $V \gtrsim 4t$ as in our analytical calculations, figure~\ref{fig2}. This suggests that   the spin-fluctuation pairing induced by kinematical interaction is the most probable  mechanism of superconductivity in the Hubbard model in the limit of strong correlations.

\newpage
\ukrainianpart

\title{Мікроскопічна теорія високотемпературної надпровідності в
сильноскорельованих електронних системах} 
%\footnote{Стаття приурочена до 60-річчя А. Швайки}}
%
\author{М. М. Плакіда}
\address{
 Об'єднаний інститут ядерних досліджень, 141980 Дубна, Росія 
}

\makeukrtitle

\begin{abstract}
Представлено послідовну мікроскопічну теорію надпровідності для сильноскорельованих електронних систем. Виведено рівняння Дайсона для нормальної та аномальної функцій Ґріна на мові проекційних (габардівських) електронних операторів. Для порівняння різних механізмів спарювання, було розглянуто узагальнену модель Габарда, де враховано міжвузлове кулонівське відштовхування і електрон-фононну взаємодію. Ми отримали $d$-хвильове спарювання з високою $T_c$, що зумовлено сильною кінетичною взаємодією електронів зі спіновими флуктуаціями, тоді як кулонівське відштовхування та електрон-фононна взаємодія при $d$-хвильовому спарюванні послаблюються. Ці результати узгоджуються з спін-флуктуаційним механізмом високотемпературної надпровідності в купратах, що раніше був запропонований у феноменологічних моделях.
\keywords сильноскорельовані електронні системи, Габардова модель, незвичайна надпровідність, купратні надпровідники
%
%\pacs 71.27.+a , 71.10.Fd , 74.20.Mn , 74.72.-h
\end{abstract}
\lastpage
\end{document}